\documentclass{appolb}
\usepackage{graphicx}

\usepackage{cite}

\usepackage{color}

\usepackage[utf8]{inputenc}

\usepackage{amsmath}
\usepackage{amssymb}

\begin{document}
\title{Features of the QCD phase diagram \\ from small, noisy, fluctuating systems%
\thanks{Presented at Excited QCD 2017}%
}
\author{Eduardo S. Fraga, Maur\' icio Hippert
\address{Instituto de F\' isica, Universidade Federal do Rio de Janeiro,\\
Caixa Postal 68528, 21941-972, Rio de Janeiro, RJ, Brazil}
}
\maketitle
\begin{abstract}
Current heavy-ion collision experiments might lead to the discovery of a first-order chiral symmetry breaking phase-transition line, ending in a second-order critical point. Nevertheless, the extraction of information about the equilibrium thermodynamic properties of baryonic matter from the highly dynamic, small, noisy and fluctuating environment formed in such collisions is an extremely challenging task. We address some of the limitations present in the experimental search for the QCD critical point. 
\end{abstract}
  
\section{Introduction}


The discovery or exclusion of an expected QCD phase transition would be a milestone in the study of the fundamental forces of nature, paving the way towards 
a better understanding of extremely dense matter.  
At high densities, this transition is expected to become of first order at a critical endpoint of unknown location. 
While the first-order transition line  would be signalled by a two-peak statistical distribution of observables and by structure formation 
from nucleation or spinodal decomposition, the second-order endpoint is characterized by long-wavelength fluctuations and universal 
scaling laws. 
The mapping of the phase diagram of strong interactions, and more specifically the search for the QCD phase transition, 
is the major goal of both the RHIC Beam-Energy Scan and new facilities aiming at producing dense baryonic matter from the collision of nuclei \cite{Xu:2016mqs}.   


A variety of experimental signatures of the QCD critical point has been discussed in the literature, mostly relying on the enhancement of 
long-wavelength fluctuations and universal behavior. 
Susceptibilities connected to conserved charges, for instance, are expected to undergo largely non-Gaussian, non-monotonic behavior, in the vicinity of the critical point \cite{Stephanov:1999zu,Stephanov:2008qz}. 
%
However, the systems which are formed in heavy-ion collisions are small, short-lived, noisy and fast evolving and are only indirectly probed. 
It is thus important to develop robust signatures of criticality, understanding how different experimental limitations affect them.

\section{Imprints of the critical point}

The proximity of a second-order critical point is marked by a divergent increase of the correlation length $\xi$. 
The associated long-range behavior makes microscopic scales irrelevant, resulting in scalings of thermodynamic susceptibilities/statistical cumulants, 
which diverge with given powers of $\xi$, determined by universal critical exponents. 


We can describe the long-range fluctuations of the order parameter $\sigma$ by a probability distribution: 
\begin{equation}
 \mathcal{P}[\sigma] \sim e^{-\Omega[\sigma]/T} \approx \displaystyle  e^{ \int d^3 x \; \left\{ \frac{(\nabla \sigma)^2}{2} + \frac{m^2_\sigma}{2} \sigma^2 {\color{red}+ \frac{\lambda_3}{3} \sigma^3 +   \frac{\lambda_4}{4} \sigma^4} + \cdots \right\}/T}
\,,
\label{eqPsigma2}
\end{equation}
where fluctuations were assumed to be of small amplitude.
In that case, we can make a Gaussian approximation by considering only the mass term, in which $m_\sigma \sim \xi^{-1}$. 
We also assume fluctuations to be homogeneous and use $\sigma_0 = \int d^3x \, \sigma(x)/V$.
Finally, we consider their couplings to observable particles via mass corrections:
\begin{equation}
 \mathcal{L}_{int} = -G\,\sigma_0 \, \vec \pi \cdot \vec \pi - g \,\sigma_0\, \bar\psi_p\, \psi_p\,,
\end{equation}
where we consider couplings to pions and protons \cite{Stephanov:2008qz}. The pion-sigma coupling can be roughly estimated to be 
around $G\sim 300$ MeV \cite{Stephanov:1999zu}.

Fluctuations of the order parameter are then coupled to observable particles and will have an impact, for instance, in fluctuations of 
particle multiplicities. 
The effects of these fluctuations can be calculated by looking at the modification of the single-particle energy levels, due 
to fluctuations of the order parameter: 
\begin{equation}
  \omega  =  \sqrt{p^2 + m_0^2 + \delta m^2 } 
       \approx   \omega_0\left[1+\dfrac{1}{2} \dfrac{\delta m^2}{\omega_0^2} - \dfrac{1}{8} \dfrac{(\delta m^2)^2}{\omega_0^4} + \cdots \right]
\,,
\label{dwds}
\end{equation}
where we have used a Taylor expansion over the mass corrections $\delta m$ from fluctuations of the order parameter. 
Expanding quantities in powers of the shift in the single-particle energies $\delta \omega_{\vec p}$ and taking averages over the fluctuations of $\sigma_0$, denoted by $\overline{(\cdots)}$, it is possible to 
calculate critical contributions to averages and correlations. 
For instance,
\begin{multline}
 \overline{\langle Q \rangle} = \langle Q \rangle_0 + \displaystyle \sum_{\vec p} \dfrac{\partial\;}{\partial \omega_{\vec p}}\langle \Delta Q \rangle_0\;\overline{\delta \omega_{\vec p}}  + 
\dfrac{1}{2} \,  \sum_{\vec p,\vec p^\prime} \dfrac{\partial\;}{\partial \omega_{\vec p}}  \dfrac{\partial\;}{\partial \omega_{\vec p^\prime}} \langle Q_1\rangle_0 \;\overline{\delta \omega_{\vec p} \, \delta \omega_{\vec p^\prime}}\,,
\label{eq-crit-Q}
\end{multline}
where $\langle \cdots \rangle_0$ denotes the usual equilibrium averages in a grand canonical ensemble and $Q$ is a generic quantity \cite{Hippert:2017xoj}. 



Near a second-order phase transition, the equilibration timescale of the system also diverges with some power of $\xi$, in a phenomenon known as critical slowing-down. 
This effect limits the growth of $\xi$ and, consequently, of possible signatures which scale with $\xi$ to some power. 
It is implemented in the ansatz equation \cite{Berdnikov:1999ph,Hippert:2015rwa}
\begin{equation}
 \dfrac{{d} \xi}{{  d} t} = A\; \left(\dfrac{\xi}{\xi_0} \right)^{2-z}\,\left(\dfrac{\xi_0}{\xi} - \dfrac{\xi_0}{\xi_{eq}(t)}\right)\, ,
\label{modBerdnikov1}
\end{equation}
where $\xi_{eq}(t) = \xi_0 \;|{t}/{\tau} |^{-\nu/\beta \delta}$, $\xi_0 \sim 1.6$ fm fixes the initial correlation length at proper time $t=-\tau$ and $\tau$ is the 
typical cooling time before reaching the neighborhood of the critical point. 
The critical exponents are given by $\alpha = 0.11$, $\nu=0.63$, $z=2 + \alpha/\nu$, $\beta=0.326$, $\delta=4.80$, coming from universality class arguments 
\cite{Guida:1996ep, Hohenberg:1977ym}. 
The parameter $A$ in Eq. (\ref{modBerdnikov1}) can be constrained by imposing causality (i.e. $d\xi/dt \leq 1$), constraining $\xi/\xi_0$ to below 
$1.3$ for $\tau = 1$ fm and below $1.9$ for $\tau = 5.5$ fm and significantly restraining signatures of criticality \cite{Hippert:2017xoj}.

\section{Effects from the experimental context}

The statistics to be measured in collision experiments are not quite the same as the ones in Eq. (\ref{eq-crit-Q}). 
They are contaminated by spurious fluctuations, modified by acceptance and efficiency limitations and are not calculated over 
direct particles only.  These effects can be introduced into our calculations in a simple fashion. 
Other effects, such as the ones from the dynamical expansion of the system are, for now, neglected.

\subsection{Acceptance window and resonance decay}


Effects from a limited acceptance window can be implemented in the calculation of multiplicity fluctuations by 
considering an acceptance probability factor $F(p)$, such that each produced particle of momentum $p$ (in modulus) 
has a probability  $F(p)$ of being detected \cite{Hippert:2017xoj}.  
For instance, if $n_p$ is the number of particles with momentum $p$, these kinematic cuts modify 
$\langle (\Delta n_p)^2 \rangle$ according to 
$
 \langle (\Delta n_p)^2 \rangle_{acc} = F(p)^2\,\langle (\Delta n_p)^2 \rangle + F(p) \big(1-F(p)\big)\, \langle n_p \rangle \,.
$



\begin{figure}[htb]
\centerline{
\includegraphics[width=12.5cm]{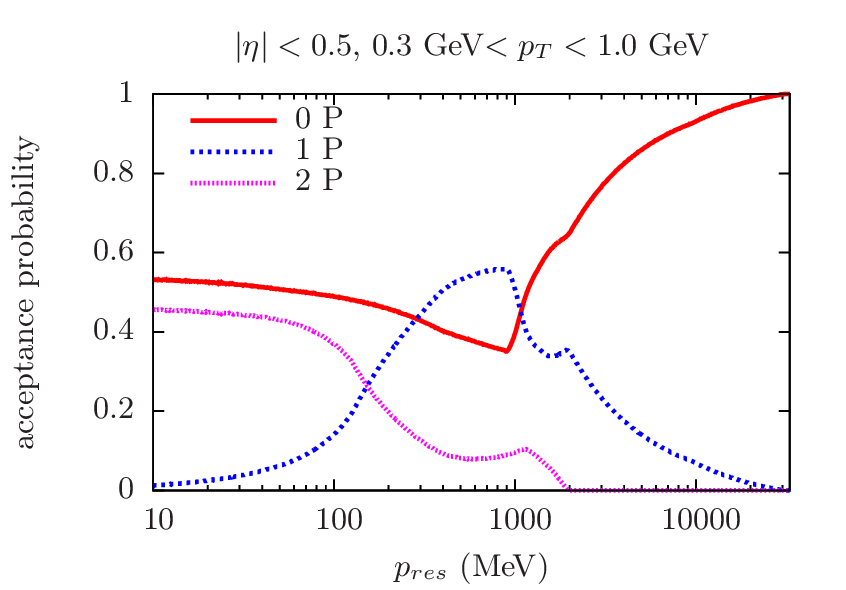}}
\caption{Probability of accepting one, both or none of the pions coming from the decay of a rho-meson, 
as a function of the total momentum, in modulus. }
\label{Fig:Res}
\end{figure}

Effects from resonance decay can be introduced in a similar fashion. 
For a decay into two particles, we consider the probabilities that one, both or neither of the 
particles produced in a single decay are found in the acceptance window. 
These probabilities, denoted as $P_1$, $P_2$ and $P_0$, respectively, are displayed in Fig. \ref{Fig:Res}, 
in the case of the decay of a rho-meson into two pions, under the cuts $0.3\, {\rm GeV} <p_T< 1.0\, {\rm GeV} $ and  $|\eta|<0.5$, 
where $p_T$ is the transverse momentum and $\eta$ is the pseudo-rapidity of the decay products. 
Results are shown as a function of 
the momentum $p$ of the resonance and 
 were calculated by using phase-space volume as a measure of probability.


\subsection{Spurious fluctuations}

Finally, spurious fluctuations coming from the imperfect control of the freeze-out thermodynamic variables, 
such as temperature, chemical potential and volume can also be included by shifting the one-particle energy levels $\omega_{\vec p}$. 
Considering spherically symmetric boundary conditions, for instance, momentum levels are distributed as $p_i = \alpha_i/R$, where $R$ is the system radius. 
These means that a geometric fluctuation of the radius of $\delta R$ will affect the energy levels through 
\begin{align}
 p_i= & \dfrac{\alpha_{i}}{R +\delta R} \approx p_{0\, i} \left[ 1 - \dfrac{\delta R}{R} + \left(\dfrac{\delta R}{R}\right)^2 + \cdots\right] \,.
\end{align}
Fluctuations of temperature and chemical potential can likewise be included by introducing the effective energy shift $\delta\omega_{T,\mu}$, such that 
$
  {\omega + \delta \omega_{T,\mu}-\mu}/{T} = {\omega-(\mu + \delta\mu)}/{T+\delta T}. 
$

In the results of Fig. \ref{Fig:pions}, we implemented temperature fluctuations of $5\%$ width and geometric fluctuations coming from the centrality bin width for the $5\%$ most central collisions.

\section{Results and final remarks}

The results above were used to calculate the average multiplicity of charged pions, $M_{\pi_{ch}}$, and its variance, $V_{\pi_{ch}}$, as a function of $\xi$. 
An acceptance window of $0.3\, {\rm GeV} <p_T< 1.0\, {\rm GeV} $ and  $|\eta|<0.5$ was used. 

\begin{figure}[htb]
\centerline{
\includegraphics[width=12.5cm]{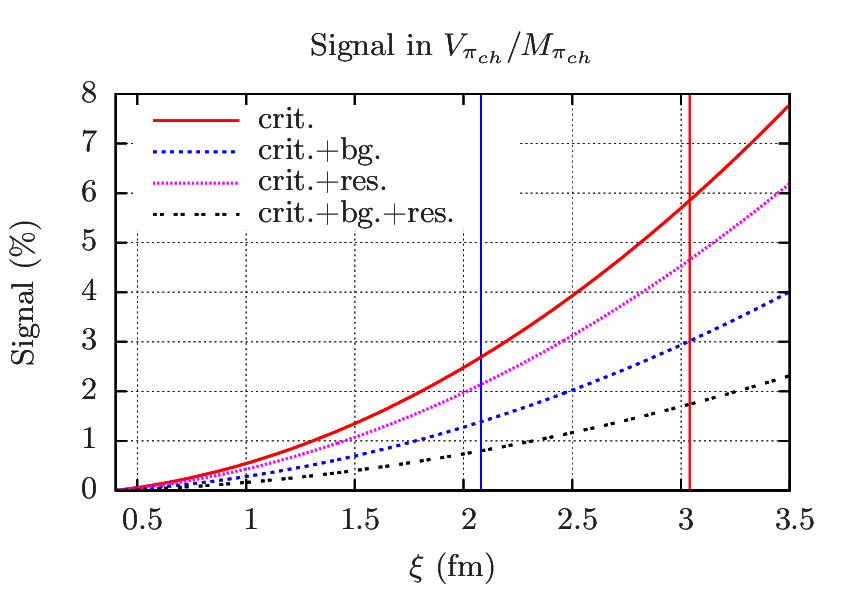}}
\caption{Signal in the variance over average ratio of the multiplicity of charged pions, as a function of $\xi$, for $0.3\, {\rm GeV} <p_T< 1.0\, {\rm GeV} $ and  $|\eta|<0.5$. 
The blue (red) line represents the limit in $\xi$ coming from $\tau = 1$ fm ($5.5$ fm).}
\label{Fig:pions}
\end{figure}

Fig. \ref{Fig:pions} displays results for the percentage by which the example-signature  $V_{\pi_{ch}}/M_{\pi_{ch}}$  grows with respect to $\xi$, 
with respect to its value at $\xi = 0.4$ fm, when only critical, background and the decay of rho-mesons are taken into account. 
More details and results can be found in \cite{Hippert:2017xoj}, where caveats are also discussed. 
Future work will extend this results to the more interesting signatures connected to protons and higher-order moments of particle multiplicities. 

We thank CNPq and FAPERJ for financial support. 



\end{document}